\documentclass[fleqn,usenatbib]{mnras}
\usepackage{natbib}
\usepackage{url}
\usepackage{rotating}
\usepackage{amsmath}
\usepackage{amsfonts}
\usepackage{amssymb}
\usepackage{graphicx}
\usepackage{float}
\usepackage{color}
\usepackage{ulem}

\def\ciii{C\,{\sc iii]}}
\def\civ{C\,{\sc iv}}
\def\mgii{Mg\,{\sc ii}}
\def\feii{Fe\,{\sc ii}}

\newcommand\lsim{\mathrel{\rlap{\lower4pt\hbox{\hskip1pt$\sim$}}
        \raise1pt\hbox{$<$}}}
\newcommand\gsim{\mathrel{\rlap{\lower4pt\hbox{\hskip1pt$\sim$}}
        \raise1pt\hbox{$>$}}}

\usepackage{hyperref}
\hypersetup{
    pdfnewwindow=true,      
    colorlinks=true,       
    linkcolor=blue,          
    citecolor=blue,        
    filecolor=blue,      
    urlcolor=blue           
}

\begin{document}

\title[Testing DB hypothesis for BBH candidates]{Testing the relativistic 
    Doppler boost hypothesis for supermassive binary black holes candidates 
    via broad emission line profiles}

\author[Song et al.]{Zihao Song$^{1,2}$, 
    Junqiang Ge$^{1}$, 
    Youjun Lu$^{1,2,\dagger}$, 
    and Xiang Ji$^{1,2}$\\
    $^1$National Astronomical Observatories, Chinese Academy of Sciences, 
    20A Datun Road, Beijing, 100101, China; luyj@nao.cas.cn  \\
    $^2$School of Astronomy and Space Science, University of Chinese Academy of 
    Sciences, No. 19A Yuquan Road, Beijing, 100049, China}

\maketitle

\begin{abstract}
Optical periodicity QSOs found by transient surveys are suggested to be 
sub-parsec supermassive binary black holes (BBHs). An intriguing interpretation
for the periodicity of some of those QSOs is that the continuum is radiated
from the accretion disk associated with the BBH secondary component
and modulated by the periodical rotation of the secondary
via Doppler-boost effect. Close to edge-on orbital orientation 
can lead to more significant Doppler-boost effect and thus are preferred 
for these systems, which is distinct from those normal type-1 QSOs with 
more or less face-on orientations. Therefore, the profiles of broad lines
emitted from these Doppler-modulated systems may be significantly different 
from other systems that are not Doppler-modulated. 
We investigate the properties of the broad emission lines of 
optical-periodicity QSOs, including both a sample of QSOs that can be 
interpreted by the Doppler-modulated effects and a sample that cannot. 
We find that there is no obvious difference in the profiles and other 
properties of various (stacked) broad emission lines of these two samples, 
though a simple broad line region model would suggest significant differences. 
Our finding raises a challenge to the Doppler boost hypothesis 
for some of those BBHs candidates with optical periodicity.
\end{abstract}


\section{Introduction}
\label{sec: Intro}

Supermassive binary black holes (SMBBHs) have long been anticipated to exist in  
centers of many galaxies as the natural products of hierarchical galaxy mergers 
\citep[e.g.,][]{1980Natur.287..307B, Yu2002} because most galaxies host a massive 
black hole (MBH) in their centers \citep[e.g.,][]{Magorrian98, 2002ApJ...574..740T, KL13}. 
Searching for sub-parsec SMBBHs has become a hot topic in recent years, and a 
number of candidates have been found according to various plausible SMBBH 
signatures, including particular line features, such as double-peaked/asymmetric 
line profile or offset lines \citep[e.g.,][]{Tsalmantza2011, Eracleous2012, 
Ju2013, Liu2014, 2019MNRAS.482.3288G}, optical-UV flux deficit 
\citep[e.g.,][]{Yan2015, Zheng16}, and periodical variation of light curves 
\citep[e.g.,][]{Graham15Nat, 2015MNRAS.453.1562G, 2016MNRAS.463.2145C, 
2018MNRAS.476.4617C, 2019ApJS..241...33L}. 
However, it is still difficult to confirm that any of these candidates is really 
a sub-parsec SMBBH system.

Among those candidates, more than $100$ are the optical periodicity QSOs
(typically with period about $1$ to a few years) found through transient 
surveys (e.g., Catalina Real-time Transient Survey, Palomar Transient Factory, 
and Rapid Response System (Pan-STARRS)), of which some were suggested to 
be due to orbital modulated accretion rate variation 
\citep[e.g.,][]{Graham15Nat,2016MNRAS.463.2145C,2019ApJ...884...36L}
or relativistic Doppler boosted continuum 
radiation from active SMBBH systems. The most intriguing example is PG\,1302-102, 
of which the optical-UV continuum variation appears to be consistent with the 
Doppler boosting of the continuum radiation from the disk associated with the 
secondary MBH \citep[e.g.,][]{2015Natur.525..351D,2019arXiv190711246X}. 
\citet{2018MNRAS.476.4617C} have 
further investigated the viability of Doppler boost interpretation for a sample 
of those SMBBH candidates with optical periodicity by using their optical-UV 
continua, the steeper the UV continua, the more significant the Doppler boost 
effect. They found that about one-third of them can be interpreted as the 
Doppler-modulated SMBBHs while others cannot, 
although these targets may be still contaminated by normal
QSOs (with a maximum fraction of 37\% in the FUV band) due to their color-dependent 
variability.

Apparently the Doppler boost effect is also dependent on the orientation of 
orbital plane relative to the line of sight (LOS), and the effects for those 
systems with close to edge-on orbital 
orientation are much more significant than that for face-on ones, 
which require higher rotating velocities to present similar Doppler boost 
effect as the edge-on cases. Therefore, the orbital orientation for those 
systems that can be interpreted by the Doppler boost effect may be more likely 
to have close to edge-on orbital orientations. According to \citet{Graham15Nat}, 
for example, the orbital inclination of PG 1302-102, defined as the angle 
between the disk normal and the LOS, requires to be $\ga 60^\circ$ if its 
optical periodicity is explained by the Doppler boost effect. However, 
normal QSOs or those QSOs with optical-periodicity but not due to the Doppler 
boost are supposed to have more face-on oriented disk according to the standard 
AGN unification model \citep[e.g.,][]{ Antonucci93, Krolik99}. 
Assuming the orbital plane is aligned with the disk, then the profiles of 
broad lines of the Doppler-modulated systems are expected to be systematically 
different from  that of those systems with insignificant Doppler boost effects 
due to the possible systematic differences in the disk orientations, since 
broad line regions (BLRs) are generally flattened and not spherical 
\citep[][]{2012ApJ...754...49P, 2014MNRAS.445.3055P, 2014MNRAS.445.3073P, 2018ApJ...866...75W}. 
If the BLR is not aligned with the BBH orbital
plane or they are significantly mis-aligned, the systematic difference may become weaker.

In this paper, we investigate the properties, especially profiles, of the 
broad emission lines of those optical-periodicity QSOs, and check whether 
there are systematic differences in the broad lines emitted from those BBH 
candidates that can be interpreted by the Doppler-modulated effects and those 
that cannot.The paper is organized as follow. In Section
~\ref{section:Preparation}, we introduce the optical-periodicity QSO samples, 
data analysis of spectroscopic data adopted from \citet{2018MNRAS.476.4617C}, 
and a BLR model to fit broad line profile. In Section~\ref{section: Result}, we 
present main results on the line properties of individual sample objects and the 
stacked line profiles for both the sample that can be interpreted by the Doppler 
boost effect and that cannot, then we test the the Doppler boost hypothesis, by 
comparing the expected stacked broad line profile, obtained for the proposed 
Doppler boost sample objects via the simple BLR model, with the observational 
one. Conclusions are summarized in Section~\ref{sec:con}. 

\section{Sample, data analysis, and a simple BLR model for line profile}
\label{section:Preparation}

\subsection{Sample}

\cite{2018MNRAS.476.4617C} collected a sample of BBH candidates via optical
periodic variations, including 21 objects of which the periodicity was 
proposed to be due to relativistic Doppler boost (DB; here after denoted as 
DB objects), and 47 objects of which the periodicity is not due to 
relativistic Doppler boost (hereafter denoted as non-DB objects). By 
cross-matching those objects with the SDSS spectral catalog DR13 
\citep{2017ApJS..233...25A}, we find that all the 21 DB objects and 40 out 
of the 47 non-DB objects have SDSS spectra. Hereafter we take those 21 DB 
objects as the DB sample and those 40 non-DB objects as the non-DB sample.

\begin{figure}
\begin{center}
\includegraphics[width=1.0\linewidth]{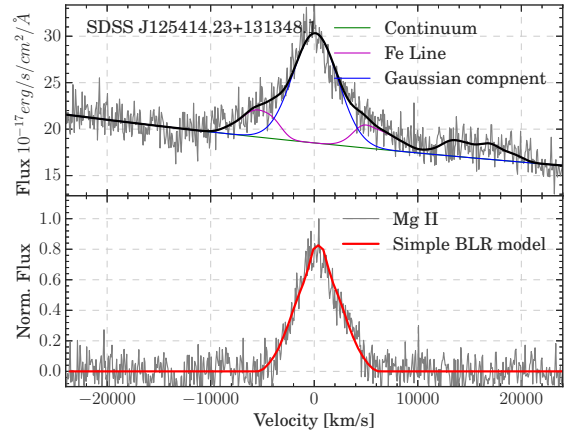}
\caption{Examples of the spectral fitting (top panel) and the 
model fitting to the line profile of \mgii\ (bottom panel; for object 
SDSS J125414.23+131348.1). In the top panel, the observed spectrum 
(grey curve) is fitted by the combination of three components, i.e., 
a power-law continuum (green), a Gaussian \mgii\ broad line (blue), 
and multiple lines from \feii\ templates by \citet{2004ApJ...611...81S} 
described by Gaussian profiles with the same velocity dispersion (magenta). 
The best fit is shown by the black curve in the top panel. In the bottom 
panel, the grey curve shows the \mgii\ profile obtained by subtracting
the best-fit continuum and \feii\ lines from the observed spectrum. 
The red line represents the best-fit \mgii\ profile to the grey curve 
obtained from the simple BLR model introduced in Section~\ref{subsec:model}.}
\label{Fig:modelfiteg}
\end{center}
\end{figure}

\subsection{Data analysis}

For the DB and non-DB samples, we can identify broad emission lines 
like \ciii, \civ, \mgii, and others from their SDSS spectra. The numbers of
these BBH candidates that have \mgii, \ciii, and \civ\ lines in their
SDSS spectra are among the top three, hence we take them for following analysis. 
We adopt a model with three components to fit each observed spectrum, i.e., 
a power-law for continuum, a Gaussian profile for each broad line, and multiple
lines from \feii\ templates \citep{2003ApJS..145...15S, 2004ApJ...611...81S} 
described by Gaussian profiles with the same velocity dispersion (which can
be different from that for the broad line). \feii\ lines are considered here
as they contaminate the above three lines (especially \mgii) significantly. 
The wavelength windows in the rest frame for the fittings to each broad line 
are $1450$-$1650$, $1800$-$2000$, and $2700$-$3000$\AA, respectively, 
and the windows for \feii\ lines are $1450$-$1530$, $1570$-$1650$, 
$1800$-$1880$, $1940$- $2000$, $2700$-$2770$, and $2830$-$3000$\AA.
A broad emission line, either \ciii, \civ, or \mgii\, is confirmed if the best
fit gives an FWHM $>1000$\,km/s and a peak flux $>3\sigma$ significance. 
Figure \ref{Fig:modelfiteg} shows an example for such a model fitting to \mgii\ 
line of an object SDSS J125414.23+131348.1 (top panel). The total numbers 
of those objects that have confirmed broad emission line \ciii, \civ, and 
\mgii\ in each sample are listed in Table \ref{table:BL}, respectively. 
As seen from Table~\ref{table:BL}, most sample objects have confirmed \mgii, 
while only a small fraction (or $\la 50\%$) of the sample objects have 
confirmed \civ\ or \ciii\ lines.

\begin{table}
\caption{Number of BBH candidates that have different broad lines in their
SDSS spectra}
\centering
\begin{tabular}{ccccc}
\hline \hline
 sample & \civ & \ciii & \mgii  \\ \hline
 DB     &  7   & 12    & 21  \\
 non-DB &  7   & 13    & 36  \\ \hline
\end{tabular}
\label{table:BL}
\end{table}

We analyze these three broad emission lines in the following ways in order to
investigate the properties of the DB and non-DB samples. First,
we stack the observed profiles of each line together for each sample, 
in which the integrated flux of each line is normalized to $1$ before stacking. 
In this way, we may find whether there is a systematic difference between the
mean line properties of the two samples. Second, we compare the FWHM 
distributions of each broad line resulting from the above model fittings to
the objects in both samples. If the DB and non-DB samples are correctly 
classified based on the relativistic DB, then this DB effect should be reflected 
by the difference between the stacked line profiles of the DB sample and that
of the non-DB sample, as well as the differences between the FWHM distributions 
of the lines for objects in the DB sample and that in the non-DB sample,
provided that the geometry of BLR is flattened and the DB 
objects are preferably viewed at a directly close to edge-on.

Beside the mean profile and the FWHM distribution, another important quantity to 
indicate the difference is the inclination angle of the BLR, 
which can not be derived without BLR modeling. To have a further understanding 
on line properties for objects in those two samples, below we introduce a 
simple BLR model in order to give a rough estimation of the inclination angle 
for each object.

\begin{figure*}
\includegraphics[width=1\linewidth]{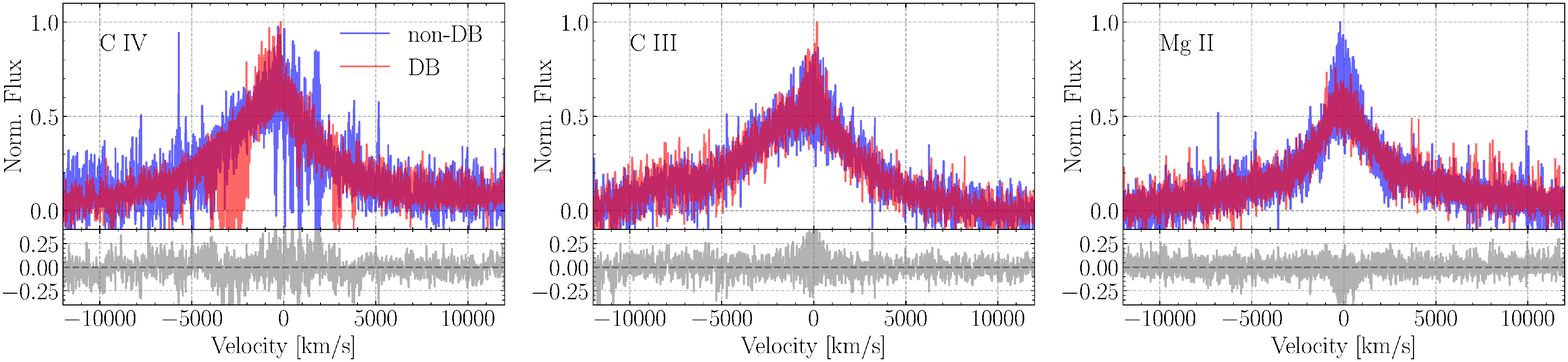}
\includegraphics[width=1\linewidth]{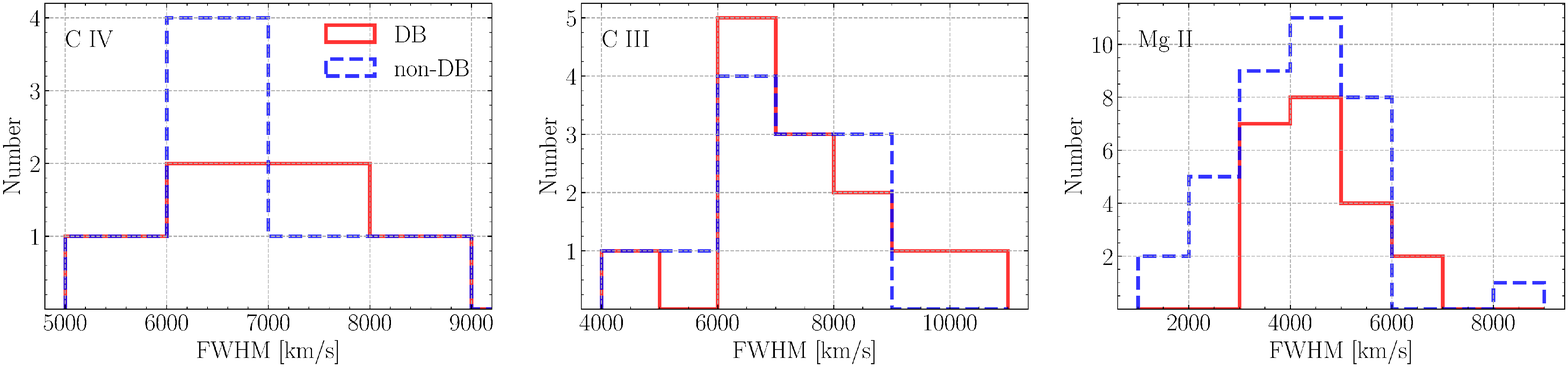}
\caption{The stacked \civ+\feii, \ciii+\feii, and \mgii+\feii\ broad lines 
(top panels from left to right) and the FWHM distributions for the \civ, \ciii, 
and \mgii\ broad lines among the objects in the DB sample and the
no-DB sample (bottom panels from left to right), respectively. 
In each top (bottom) panel, red and blue curves (histograms) represent
the results obtained for the DB and non-DB sample, respectively. 
In each of the three small panels right below the top panels, the grey
curve shows the difference between the stacked line profile of the DB
sample (red curve) and that of the non-DB sample (blue curve).
This figure illustrates that there is no significant statistical 
difference between the profiles of the broad lines emitted from objects
in the DB sample and those from objects in the non-DB sample.
}
\label{Fig:stacked_FWHM}
\end{figure*}

\subsection{A simple model for the broad line fitting}
\label{subsec:model}

The broad emission line profiles are determined by the BLR geometry,
kinematics, and structures. We assume that BLR
is composed of a large number of clouds/clumps, rotating around the
central supermassive black hole(s) on circular orbits due to gravity. 
We assume that the number density distribution of 
the BLR clouds/clumps is only a function of radius, and it is cut at an 
inner radius of $R_{\rm in}$. The BLR may be also flattened. 
Based on these simple assumptions, the line profile can be simply 
estimated as \citep []{1975LicOB.689....1B, 1976ApJ...203..714B}
\begin{eqnarray}
F(\lambda)& \simeq & \int_{R_{\rm in}}^{R_{\rm out}} \int_{\theta_{1}}^{\theta_{2}}
\int_{0}^{2\pi}n_{\rm c}(r) j_{\rm c}(r)\delta \left[\lambda-\lambda_{0}\left(1+\frac{\upsilon_{\parallel}}{c}\right)\right] \nonumber  \\ 
& & \times r^{2} 
\sin\theta dr d\theta d\phi, 
\label{Equa:Origin}
\end{eqnarray}
where $\lambda_0$ is the center of a line, $n_{\rm c}(r)$ the number density 
distribution of clouds, $j_{\rm c}(r)$ the emissivity, $v_{\parallel}$ is the 
projected velocity of a cloud along the LOS, $\theta_1$ and $\theta_2$ defining
the flattening of the BLR. In our calculation, $j_{\rm c}$ 
is set to be constant.

The semi-major axis of a BBH system with a total mass of $M_{\bullet\bullet}$ 
and a period of $P_{\rm orb}$ can be roughly estimated as
\begin{equation}
a_{\rm BBH} = 6.75 \bigg(\frac{M_{\bullet\bullet}}{10^{8}M_{\odot}}\bigg)^{1/3} 
\left(\frac{P_{\rm orb}}{1460\,\rm day}\right)^{2/3} \mbox{lt-days}.
\end{equation}
All the sub-parsec BBH candidates in our sample have periodic variations
on timescales around a few years, which are taken as the orbital periods
of the binary systems. Adopting the above equation, we can thus estimate 
the semi-major axis for each BBH candidate in our sample by adopting those
values for $M_{\bullet\bullet}$ and $P_{\rm orb}$ given in 
\cite{2015MNRAS.453.1562G} and \cite{2016MNRAS.463.2145C}.

For an active MBH, the BLR size can be inferred from its optical luminosity 
by using the empirical relationship between BLR size and optical luminosity
given by \citep[]{2013ApJ...767..149B, 2016MNRAS.459L.124L}, i.e.,
\begin{equation}
\log (R_{\rm BLR}^L/\textrm{lt-day}) \simeq  1.527+ 0.533 \log \left[
    \frac{\lambda L_{\lambda}(5100\rm \AA)}{10^{44}{\rm erg \ s^{-1}}}\right]\,  
\label{EQ:BLR}
\end{equation}
with a scatter of $\sim 0.13-0.2$\,dex \citep[see also][]{2000ApJ...533..631K, 
2007ApJ...659..997K}.
This tight relationship is basically a result of the photon-ionization nature 
of broad emission lines \citep{1997iagn.book.....P}. For this reason, the BLR 
size of a BBH system can be similarly estimated if BLR is far away from the 
BBH, although the difference in the potential of a BBH system from that 
of a single black hole system may have some effect. We therefore adopt 
Equation~(\ref{EQ:BLR}) to estimate the BLR size for each BBH candidate.

We find that $R_{\rm BLR}^{\rm L}$ of those BBH systems are substantially 
larger than their semi-major axes according to above calculations. The 
ratio $R_{\rm BLR}^{\rm L}/a_{\rm BBH}$ ranges from about $6$ to $136$ for 
the BBH candidates in our sample, with a median value of $42$. Therefore,
it is safe to assume that those broad lines are emitted from regions 
rotating around but far away from the binary systems, i.e., 
a circum-binary BLR. 

The BLR geometry and structure have been intensively studied in the past 
several decades and it becomes clear only recently. \citet{2014MNRAS.445.3055P} 
find that the BLRs of a few AGNs are significantly flattened, with an
opening angle of $\sim 27^{\circ}-49^{\circ}$. \citet{2018Natur.563..657G} 
make a breakthrough discovery on the BLR structure of 3C\,273 by resolving 
its kinematics and find the half opening angle of the flattened BLR is about
$45^{\circ}$, consistent with the results by  \citet{2014MNRAS.445.3055P}.
Therefore, we model the geometry of a BLR similar to that described in 
Section 2.4 of \citet{2014MNRAS.445.3055P}, and define it with six free 
parameters: the mean and inner radius of BLR ($R_{\rm BLR}, R_{\rm in}$), 
radial shape parameter $\beta$, opening angle $\theta_{\rm o}$, inclination 
angle $i_{\rm BLR}$, and the central black hole mass $M_{\rm BH}$.

For each broad emission line, we perform the Markov Chain Monte Carlo (MCMC) 
calculations by using the software package ``emcee'' \citep{FM2013}
to derive the best fitting results of the above six parameters. The initial 
parameter setups are as follows. We allow $R_{\rm BLR}$ to vary from 
$0.1R_{\rm BLR}^L$ to $5R_{\rm BLR}^L$.\footnote{Note that the estimated MBH 
mass for each object may depend on the assumption of the BLR geometry and 
others and thus biased from the true MBH mass by a factor of $3$ or so 
\citep{2001ApJ...551...72K}. However, the line profile basically depends on 
the velocity field of the broad line clouds, which depends on 
$\sqrt{M_{\bullet\bullet}/d_{\rm c}}$ with $d_{\rm c} \sim R_{\rm BLR}$ 
defined as the distance from a cloud to the mass center, if assuming circular 
orbits. Therefore, the uncertainties in the mass estimates can be absorbed 
into the uncertainties of $R_{\rm BLR}$. For this reason, we assume 
$R_{\rm BLR}$ varies from $0.1R_{\rm BLR}^L$ to $5R_{\rm BLR}^L$ (by also 
considering the scatter in the estimates of $R_{\rm BLR}^{\rm L}$ according 
to Eq.~\ref{EQ:BLR}) and we do not separately consider the mass uncertainties 
in our simple model for fitting the broad line profile.}
For each $R_{\rm BLR}$, we set the inner radius of the BLR as $R_{\rm in}$ 
in the range of $[0,R_{\rm BLR}]$. The inclination angle ($i_{\rm BLR}$) is set to be in 
the range from $0^{\circ}$ (face-on) to $90^{\circ}$ (edge-on). The open angle 
varies in range of $[0^{\circ}$, $90^{\circ}$], and central black hole mass are 
in the range of $[0.1M_{\bullet\bullet},5M_{\bullet\bullet}]$. 
The bottom panel of Figure~\ref{Fig:modelfiteg} shows an example of the fitting, 
in which the model profile (red line) matches the observed line profile well
(with a reduced $\chi^2_{\nu}\sim 1$). 

\section{Results}
\label{section: Result}

We present our main results on stacked line profiles and fitting the broad 
line profiles via the simple BLR model introduced in Section~\ref{subsec:model}.

\subsection{Stacked line profiles}

Figure~\ref{Fig:stacked_FWHM} shows the stacked \civ+\feii\ (top-left panel), 
\ciii+\feii\ (top-middle panel), and \mgii+\feii\ (top-right panel) for the 
DB and non-DB samples, respectively, and also shows the FWHM distributions 
of those lines among the objects in each sample (from bottom left 
to right panels). In each top (bottom) panel, the blue and red curves 
(histograms) represent results obtained for the non-DB sample and 
DB sample, respectively. The difference between the two stacked
profiles shown in each top panel is correspondingly shown in a small panel
right below it, and it is fluctuating around $0$ at different
wavelengths. We also apply t-test to check whether the two FWHM 
distributions shown in each bottom panel are significantly different from 
each other, and we obtain t-test p-values as $0.31$ (\civ), $0.34$ (\ciii), 
and $0.38$ (\mgii), respectively, which suggest that the FWHM distribution 
of each line of the DB objects is statistically not different from that of 
the non-DB objects. This figure clearly demonstrates that the stacked profiles 
of each broad line obtained for the DB and non-DB samples are almost the same.

Under the Doppler hypothesis, however, we note here the stacked broad 
line profile for the proposed DB objects is expected to be different from 
that of non-DB objects as the proposed DB objects are supposed to be viewed
preferentially at close to edge-on direction while the non-DB objects are 
not, if assuming that the non-DB and DB objects have the same BLR geometry
and structure. This expectation seems to be in contradict with the observational
results presented above. To further understand the observational results and 
its constraints on the Doppler boost hypothesis for some BBH candidates, we
further adopt the simple BLR model to investigate the broad line properties
of the DB and non-DB objects below.

\subsection{Line profile Modelling and Observational Constraints 
on the DB hypothesis}
\label{subsec:blmodel}

We apply the Python MCMC code ``emcee'' to constrain parameters of the simple 
BLR model introduced in Section~\ref{subsec:model}, by fitting the \mgii\ 
broad line of each object in both the DB and non-DB samples. Here we only 
consider \mgii\ because it is identified in the SDSS spectra of most objects 
in our sample (Table~\ref{table:BL}), while other lines are only identified in 
the SDSS spectra of a small number of sample objects. By matching the observed 
line profile, we can derive the inclination angle and the BLR size of each BBH 
system.

\begin{figure}
\begin{center}
\includegraphics[width=1\linewidth]{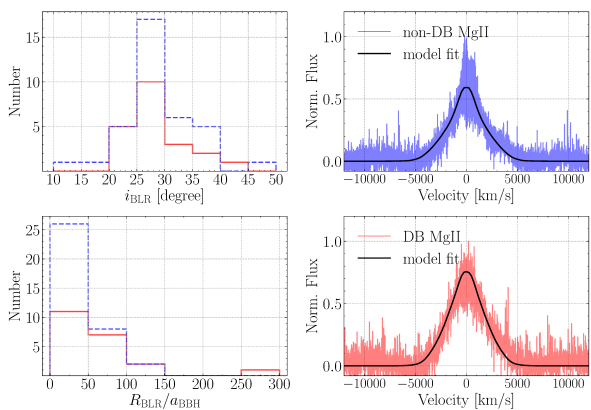}
\caption{Top- and bottom-left panels show the distributions of the best-fit 
model parameters (assuming the simple BLR model introduced in 
Section~\ref{subsec:model}), i.e., inclination angle $i$ and BLR size 
$R_{\rm BLR}$ (in unit of $a_{\rm BBH}$), among the objects in the non-DB 
(blue histogram) and DB (red histogram) samples, respectively. Top- and 
bottom-right panels show the stacked line profiles for \mgii\ (by subtracting 
the best-fit continuum and \feii\ lines) obtained for the non-DB (blue curve) 
and DB (red curve) samples, and also the stacked modelled line profile 
(black lines) by adding the best-fit \mgii\ profile of each object in the two 
samples, respectively. 
}
\label{Fig:model1}
\end{center}
\end{figure}

Figure~\ref{Fig:model1} shows the distributions of the model parameters 
[i.e., inclination angle $i$ (top-left panel) and mean BLR size in
unit of the BBH semi-major axis (bottom-left panel)] obtained from the 
simple BLR model, and the stacked profiles obtained from the best-fits 
to \mgii\ for the DB (bottom-right panel) and non-DB (top-right panel) 
objects, respectively. As seen from this figure, the inclination angles 
obtained from the model fittings to both DB and non-DB objects are mostly 
in the range from a few to $\sim 45$
degree, which suggests that those sample objects are observed at orientations 
preferentially face-on. The best-fit BLR size is significantly larger than the
semi-major axis for each BBH candidate ($R_{\rm BLR}/a_{\rm BBH}$ in the range 
from $14$ to $286$),
this again validates the assumption of circum-binary BLR made in 
Section~\ref{subsec:model}.  The stacked profiles obtained from the 
best-fits of \mgii\ lines (black lines in right panels) also match the 
observed profiles well for both the non-DB (top-right panel) and DB 
(bottom-right panel) samples. 

\begin{figure*}
\begin{center}
\includegraphics[width=1\linewidth]{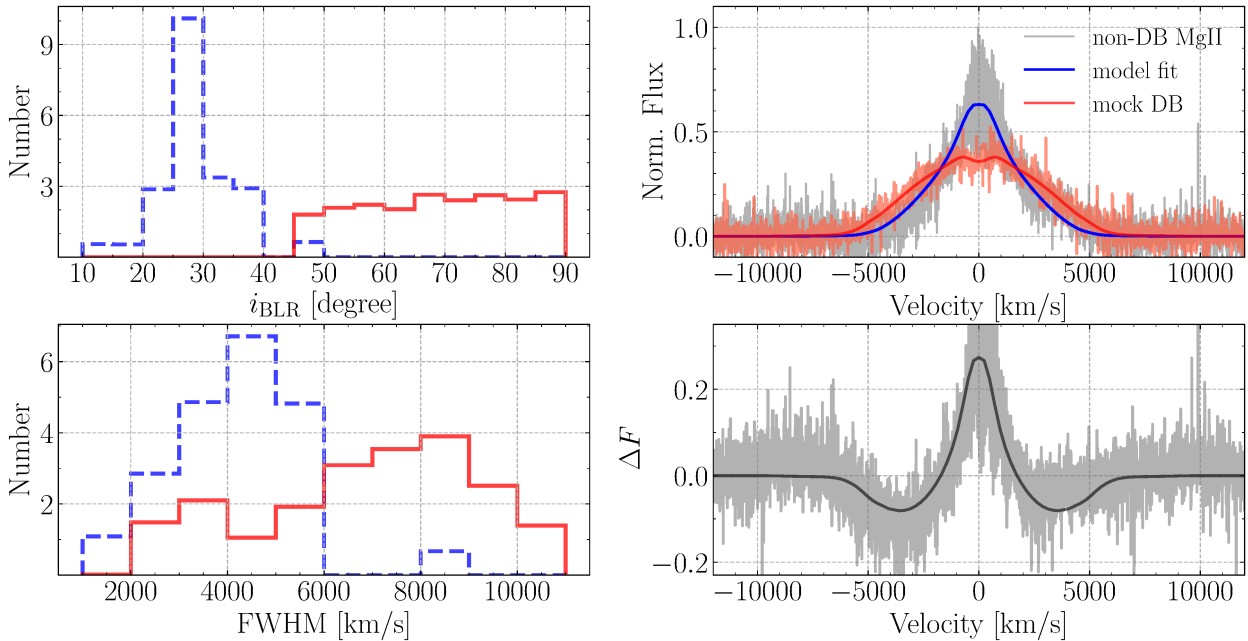}
\caption{Expected stacked \mgii\ profile of the DB sample if those objects
have flattened BLR geometry and are viewed with nearly edge-on orientations 
as suggested by the Doppler boost hypothesis to their periodicity. In the 
top-left panel, blue histogram represents the distribution of inclination 
angles for those $36$ objects in the non-DB sample obtained from the 
simple BLR model fitting, and the red histogram represents the inclination
angle distribution of $21$ mock DB objects randomly selected over the range 
from $\sin i = \sin 45^{\circ}$ to $\sin 90^{\circ}$. In the bottom-left 
panel, the blue histogram represents the FWHM distribution for those $36$ 
objects in the non-DB sample, while the red histogram represents an FWHM 
distribution of the $21$ objects in the mock DB sample randomly selected 
from the non-DB sample by replacing their inclination angles with those 
for the mock DB objects shown in the top-left panel (red histogram). 
The grey, blue, and red curves in the right panel show the stacked observed 
\mgii\ profile for the non-DB sample (same as the blue curve in the top-right 
panel of Fig.~\ref{Fig:model1}), the stacked modelled \mgii\ profile for the 
non-DB sample (blue curve; same as the black curve in the top-right panel of 
Fig.~\ref{Fig:model1}), and the expected stacked \mgii\ profile for the mock 
sample of $21$ mock DB objects (red curve with the corresponding flux error
, see Section~\ref{section: Result} for details), respectively. 
The bottom-right panel shows the profile difference between the mock DB sample 
and the non-DB sample, the valleys and summit features shown here are significantly 
different from that shown in Fig.~\ref{Fig:stacked_FWHM}.}
\label{Fig:model2}
\end{center}
\end{figure*}

The inclination angles for the proposed DB objects are expected to be 
systematically larger than those for the non-DB objects if the Doppler boost
hypothesis is correct \citep{2018MNRAS.476.4617C}, as DB objects are
preferentially observed at close to edge-on orientations. Such an expected 
differences in inclination angles would lead to significant difference between 
the stacked broad line profiles of the DB objects and those of the non-DB 
objects. To check such line profile differences, we take the best-fits of 
BLRs for \mgii\ lines of those $36$ non-DB objects as templates, and randomly 
choose $21$ of them and replace the inclination angle for each of these $21$ 
objects by a new one randomly selecting from $\sin i$ over the range from 
$\sin 45^{\circ}$ to $\sin 90^{\circ}$ [see inclination angle 
distribution (red histogram) in top-left panel]. These $21$ objects are taken 
as the mock DB sample, of which the \mgii\ profiles can be calculated 
according to the simple BLR model and added together to form a stacked \mgii\ 
line profile that is expected for the proposed DB sample under the Doppler 
boost hypothesis. 
For each mocked profile, we add the flux error to it based on the 
observed flux error of each \mgii\ broad line by assuming that the flux uncertainty 
at each wavelength follow a normal distribution. The stacked profile is then 
obtained by stacking all the mocked profiles of QSOs in the mock DB sample,
as shown by the red curve and the associated shaded region in the top-right 
panel of Figure~\ref{Fig:model2}. Obviously the expected stacked \mgii\ 
profile of the mock DB sample appears top-flat and has a weak double-peaked 
feature, different from that for the non-DB sample with a single peak, and it 
is also significantly broader than that for the non-DB sample. The detailed 
differences of stacked \mgii\ profiles between the mock DB sample and the 
non-DB sample are also shown in the bottom-right panel of Figure~\ref{Fig:model2}, 
which present significantly different valleys and summit from that shown in 
Figure~\ref{Fig:stacked_FWHM} (see also differences in the FWHM distributions 
shown in bottom-left panel of Fig.~\ref{Fig:model2}). This expected stacked 
\mgii\ profile is significantly different from that obtained directly from 
observations for the DB sample classified in \cite{2018MNRAS.476.4617C}, which 
provides a robust evidence to disapprove the DB and non-DB sample 
classifications obtained under the Doppler boosting hypothesis in 
\cite{2018MNRAS.476.4617C}, assuming that the orbital plane is aligned with 
the disk for each target, and those DB objects and non-DB objects have similar 
BLR geometry and structures.

The DB QSOs maybe contaminated by those with color-dependent variability of 
$20\%$ and $37\%$ in the nUV and fUV bands, respectively, that mimic the 
Doppler boosting signatures \citep{2018MNRAS.476.4617C}. 
In the DB sample, there are 7 QSOs identified by 
both the NUV and FUV bands, and 14 QSOs only by the NUV band. 
To clarify how the contaminated QSOs may affect our statistical analyses 
on the comparison of DB and non-DB samples, we perform Monte-Carlo simulations
to construct a new DB sample by removing randomly selected 20\% and 37\% 
of the QSOs selected from the NUV and FUV bands in the DB sample, respectively. 
With the re-constructed DB sample, we perform the same analysis procedures 
as done to the DB sample above. By performing ten thousands of such simulations,
we find that the stacked line profiles for those re-constructed samples are all similar to 
that of DB sample, without removing $20\%$ and $37\%$ sources (red curve in the 
bottom-right panel of Fig.~\ref{Fig:model1}), and no one is similar to the red curve shown in Figure~\ref{Fig:model2}.
These simulations verify that although the DB sample is possibly 
contaminated by normal QSOs, the sample properties of the DB sample are still 
statistically robust and the possible contamination dose not affect our conclusion.

\subsection{Orbital orientations of the DB sample}
\label{sec:orb}
The above analyses are based on the assumption that the BBH orbital plane is 
aligned with disk, however, whether they are really aligned or not is still 
unclear. To figure out how this alignment or misalignment can affect our 
understanding on the DB sample, we try to explore more details on the 
orientations of the orbital plane and BLR as follows. 

For a BBH system corotating in a circular orbit, the amplitude of the observed 
flux variability caused by the Doppler boost to the first order in $v/c$ 
\citep{2018MNRAS.476.4617C} is:

\begin{equation}
    A = (3-\alpha_{\nu})\frac{v}{c}\sin i_{\rm orb} ,
    \label{EQ:amplitude}
\end{equation}
where $v=\frac{1}{1+q} \left(2\pi\frac{GM_{\bullet\bullet}}{P_{\rm orb}}\right)^{1/3}$ 
is the orbital velocity, $i_{\rm orb}$ is the inclination angle of the orbital 
plane, $q$ is the mass ratio. With the observed amplitude $A_{\rm obs}$ and 
$\alpha_{\nu}$, the orbital inclination angle is only determined by the orbital 
velocity $v$, i.e., 

\begin{equation}
    \sin i_{\rm orb}  =  \frac{A_{\rm obs}}{3-\alpha_{\nu}}\frac{c}{v}
    \label{EQ:sini}   
\end{equation}    

Once the orbital period ($P_{\rm orb}$) and total mass ($M_{\bullet\bullet}$) 
are given by observations, the orbital velocity $v$ only varies with $q$. 
Since  $q$ ranges from 0 to 1, the maximum and minimum values of the orbital 
inclination angle can then be derived as

\begin{equation}
    \sin i_{\rm orb}^{\rm max} = \frac{A_{\rm obs}}{3-\alpha_{\nu}}\frac{2c}
                    {\left(2\pi\frac{GM_{\bullet\bullet}}{P_{\rm orb}}\right)^{1/3}}  \\
\end{equation}

\begin{equation}
    \sin i_{\rm orb}^{\rm min} =  \frac{A_{\rm obs}}{3-\alpha_{\nu}}\frac{c}
                    {\left(2\pi\frac{GM_{\bullet\bullet}}{P_{\rm orb}}\right)^{1/3}}
\end{equation}

After taking the observed $A_{\rm obs}$ and $\alpha_{v}$ given in Table~{2} of 
\cite{2018MNRAS.476.4617C}, the total mass $M_{\bullet\bullet}$ and periods 
$P_{\rm orb}$ of the BBH systems given by \cite{2015MNRAS.453.1562G} 
and \cite{2016MNRAS.463.2145C}, we can calculate $\sin i_{\rm orb}^{\rm max}$ 
and $\sin i_{\rm orb}^{\rm min}$ for each QSO in the DB sample. For the only 
one QSO in the DB sample (SDSS J154409.61+024040.0) that has 
$\sin i_{\rm orb}^{\rm min}>1$, we simply set $\sin i_{\rm orb}^{\rm min}=1$ 
and $\sin i_{\rm orb}^{\rm max}=1$ for further analyses.

With the calculated $i_{\rm orb}^{\rm min}$ and $i_{\rm orb}^{\rm max}$, 
we assume that $i_{\rm orb}$ is represented by the median value of 
[$i_{\rm orb}^{\rm min}$, $i_{\rm orb}^{\rm max}$] as
\begin{equation}
i_{\rm orb}=\frac{i_{\rm orb}^{\rm min}+i_{\rm orb}^{\rm max}}{2},
\label{EQ:orb_medium}
\end{equation}
and the error of each $i_{\rm orb}$ is roughly
\begin{equation}
\sigma_{i}^{\rm orb} =  \frac{i_{\rm orb}^{\rm max}-i_{\rm orb}^{\rm min}}{2}
\end{equation}

\begin{figure}
\begin{center}
\includegraphics[width=1\linewidth]{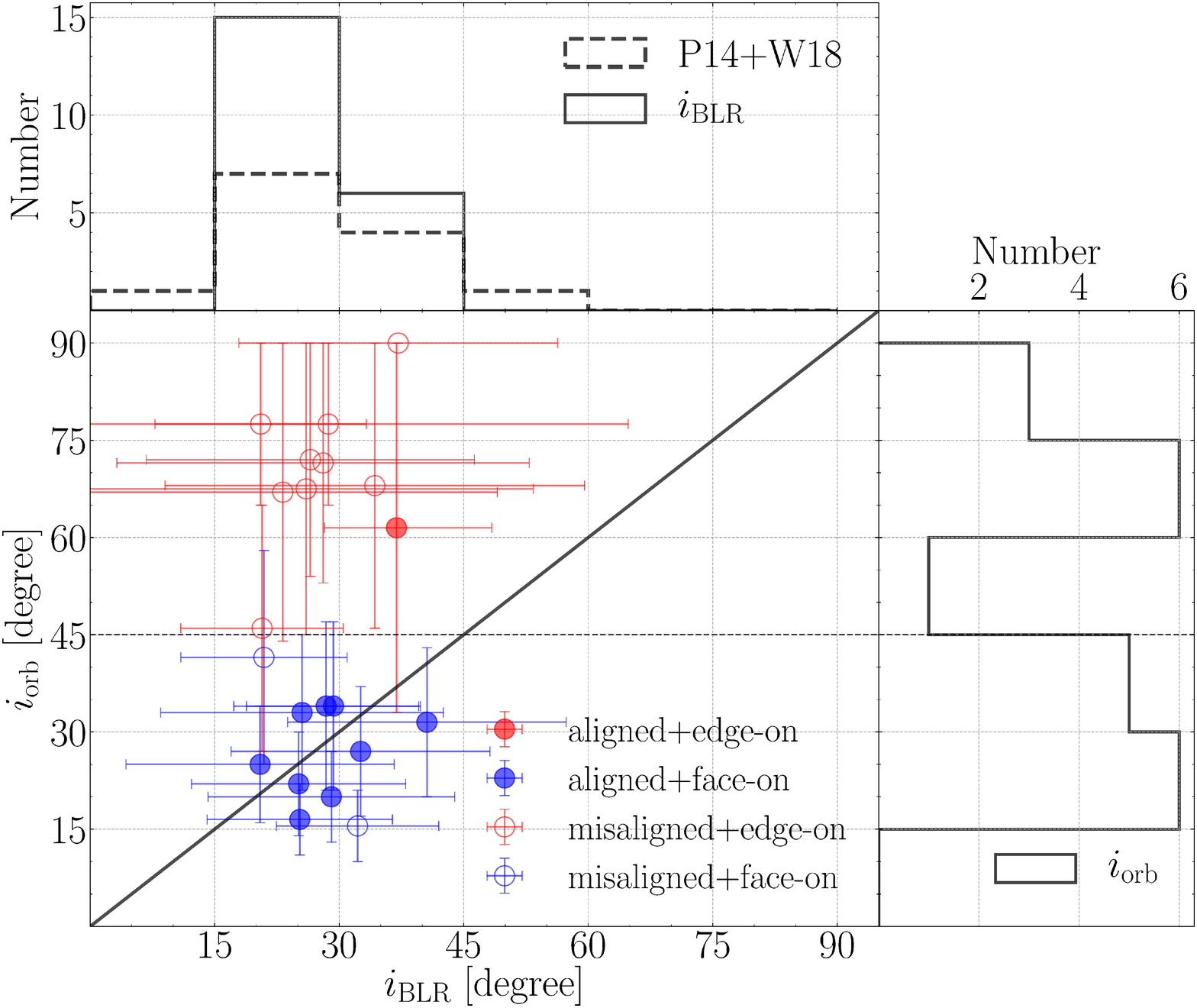}
\caption{
Distributions of $i_{\rm orb}$ and $i_{\rm BLR}$ for QSOs in the DB sample. 
The DB sample are classified into four subsamples, based on the alignment or 
misalignment of the orbital plane and BLR, and the viewing angle of SMBBH 
orbital plane, i.e., face-on or edge-on, of each QSO, i.e., 1) aligned and 
edge-on (red filled circles), 2) aligned and face-on (blue filled circles), 
3) misaligned and edge-on (red open circles), and 4) misaligned and face-on 
(blue open circles). Top panel and right panel show the distributions of 
$i_{\rm BLR}$ (black histogram) and $i_{\rm orb}$ (black histogram) of the 
DB sample, respectively. In the top panel, we plot the $i_{\rm BLR}$ 
distribution with black histogram. For comparison, the dashed histogram 
presents the $i_{\rm BLR}$ distribution of normal QSOs 
\citep[]{2014MNRAS.445.3073P, 2018ApJ...866...75W}.}

\label{Fig:incli}
\end{center}
\end{figure}

Figure~\ref{Fig:incli} shows the distributions of inclination angles of the BBH orbital 
plane ($i_{\rm orb}$) and BLR ($i_{\rm BLR}$). In the DB sample, all the QSOs have 
$i_{\rm BLR}< 45^\circ$, which are close to face on. The $i_{\rm BLR}$
distribution of the QSOs in the DB sample have no significant difference with that of 
those normal QSOs (top panel) given by \cite{2014MNRAS.445.3073P} 
and \cite{2018ApJ...866...75W}. Unlike the $i_{\rm BLR}$, those orbital inclination angles $i_{\rm orb}$ 
show a bimodal distribution (right panel), which may suggest that not all the BBH orbital 
planes are aligned with their BLRs.

Based on the difference between the two inclination angles 
($\Delta i = i_{\rm orb}-i_{\rm BLR}$) and their corresponding errors, 
we find that 10 QSOs (filled circles in Figure~\ref{Fig:incli}) 
have $i_{\rm orb}$ and $i_{\rm BLR}$ roughly the same with each other, 
which indicates that their BBH orbital plane and the BLR are aligned. For the 
other 11 QSOs (open circles in Figure \ref{Fig:incli}), their $i_{\rm orb}$ 
and $i_{\rm BLR}$ are significantly different from each other even after 
considering of their large uncertainties, which means that the BBH orbital 
planes in those systems are misaligned with the BLR planes.

According to the alignment or misalignment between the BBH orbital plane and 
BLR, and the face-on or edge-on view of BBH orbits, we further divide the DB 
sample into four subsamples: 1) 10 aligned systems (filled circles), 2) 11 
misaligned systems (open circles), 3) 11 face-on orbital plane systems 
($i_{\rm orb} \le 45^\circ$, blue circles), and 4) 10 edge-on orbital plane 
systems ($i_{\rm orb}>45^\circ$). With these classifications, we then explore 
whether their stacked broad line profiles and FWHM distributions have different 
features.

\begin{figure}
\begin{center}
\includegraphics[width=1\linewidth]{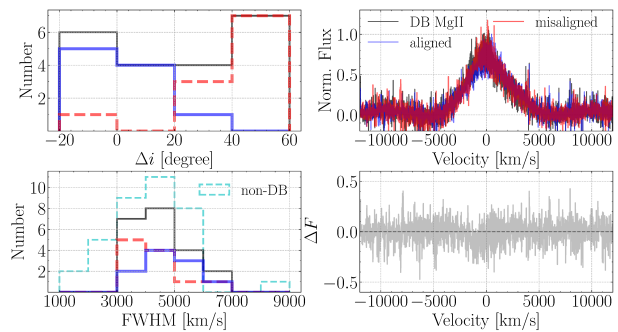}
\caption{
Stacked \mgii~profiles and FWMH distributions of QSOs in the DB sample that are 
classified by the alignment and misalignment of $i_{\rm orb}$ and $i_{\rm BLR}$. 
Top-left panel shows the inclination angle difference 
($\Delta i = i_{\rm orb}-i_{\rm BLR}$) of 10 aligned (blue histogram), 11 
misaligned (red dashed histogram), and all the 21 (black histogram) QSOs in 
the DB sample. Bottom-left and top-right panels show the corresponding 
\mgii~FWHM distributions and stacked profiles of the three subsamples labelled 
with colors the same as those in the top-left panel, respectively. For 
comparison, the FWHM distribution of the non-DB sample (cyan histogram) also 
shows in the bottom-left panel. Bottom-right panel shows the profile difference 
($\Delta F = F_{\rm aligned}-F_{\rm misaligned}$) between those two stacked 
profiles from the 10 aligned and 11 misaligned QSOs.}
\label{Fig:aligned}
\end{center}
\end{figure}

\begin{figure}
\begin{center}
\includegraphics[width=1\linewidth]{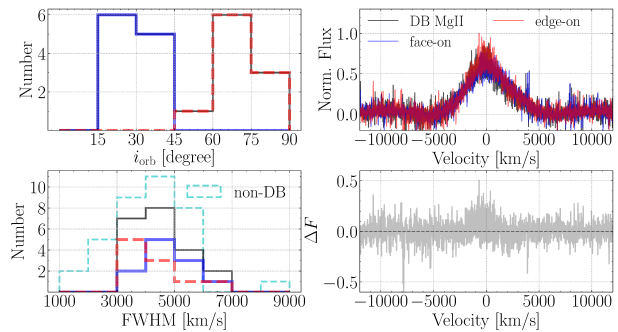}
\caption{
Stacked \mgii~profiles and FWMH distributions of QSOs in the DB sample that are 
classified by the face-on and edge-on viewing angle $i_{\rm orb}$. Top-left 
panel shows the orbital inclination angle $i_{\rm orb}$ distributions of 11 
face-on (blue color), 10 edge-on (red color), and all the 21 (black color) 
QSOs in the DB sample. Bottom-left and top-right panels show the corresponding 
\mgii~FWHM distributions and stacked profiles of the three subsamples labelled 
with colors the same as those in the top-left panel, respectively. For 
comparison, the FWHM distribution of the non-DB sample also shows in the 
bottom-left panel (cyan histogram). Bottom-right panel shows the profile 
difference ($\Delta F = F_{\rm face-on}-F_{\rm edge-on}$) between those two 
stacked profiles from the 11 face-on and 10 edge-on QSOs.}
\label{Fig:edgeon}
\end{center}
\end{figure}

Figure~\ref{Fig:aligned} shows the results of 10 aligned and 11 misaligned QSOs 
in the DB sample. Although the two kinds of QSOs have different $\Delta i$ 
distribution (top-left panel), no significant difference appears for both the 
\mgii~FWHM distributions (bottom-left panel) and stacked profiles (top-right 
panel). To check the detailed difference between the two stacked profiles, we 
show the flux residuals $\Delta F=F_{\rm aligned}-F_{\rm misaligned}$ in the 
bottom-right panel, which fluctuate around zero and have no features like the 
valleys and summit that shown in Figure~\ref{Fig:model2}.

Figure~\ref{Fig:edgeon} shows the \mgii\, FWHM distributions (bottom-left panel) 
and the stacked \mgii\, line profiles (top-right panel) for those QSOs in the 
face-on and edge-on subsamples, respectively.  Apparently there is no 
statistical difference between these two \mgii\,FWHM distributions, and the 
stacked \mgii\, profiles from the face-on and edge-on subsamples are quite 
similar with a flux residual in the velocity space close to zero as shown in 
the bottom-right panel.

Based on the above analyses, neither the aligned or misaligned nor the face-on 
or edge-on classifications of the DB sample present any distinguishable 
difference for both FWHM distributions and stacked \mgii\ broad lines. This 
conflicts with what we expect to the DB sample: the orbital orientations of 
these BBH systems distribute from face-on to edge-on, even though there is no 
correlations between $i_{\rm orb}$ and $i_{\rm BLR}$, we should still have 
opportunities to observe the BLR in edge-on like inclination angles for a DB 
sample with 21 QSOs. Current results support that we can observe those BBH 
orbits in both edge-on and face-on, but all of them only have face-on oriented 
BLRs, which further enhance the challenge for using the Doppler boost 
hypothesis to interpret the QSOs with optical periodicity in the DB sample.

\section{Conclusions}
\label{sec:con}

Optical periodicity QSOs are proposed to be candidates of the sub-parsec 
BBHs. The periodic variations of a number of these candidates were suggested
to be due to orbital modulated Doppler boosted continuum radiation from the 
secondary component of the BBH systems if viewing with close to edge-on 
orientations \citep{2015Natur.525..351D, 2018MNRAS.476.4617C}. With this 
relativistic Doppler boost hypothesis, \cite{2018MNRAS.476.4617C} further 
divided these BBH candidates into two classes according to their spectral 
properties, i.e., one class that can be explained by the Doppler boost (DB) 
effect (DB objects) and the other class that cannot (non-DB objects).

In this paper, we investigate the properties of the broad emission lines of
both the proposed DB and non-DB objects by using their SDSS spectra and 
further test the DB hypothesis. We find that the properties of various broad
emission lines of the DB objects are similar to those of the non-DB objects
and the stacked \civ, \ciii, and \mgii\ line profiles of the DB sample are 
almost the same as those of the non-DB sample. Under the DB hypothesis, however, 
we demonstrate that the stacked broad lines of the DB sample, such as \civ, 
\ciii, \mgii\, are expected to be double-peaked and significantly broader than
those of the non-DB sample, by reasonably assuming that the DB objects have 
similar flattened BLR geometry and structures as that of the non-DB objects 
but viewing preferentially at close to edge-on orientation under the DB 
hypothesis. This expectation is in contradiction with the observational
results of no difference between the stacked broad emission line profiles 
of the proposed DB and non-DB samples, which raises a significant challenge
to the Doppler boost interpretation for some of those BBHs candidates 
suggested by their optical periodicity.

By deriving the inclination angles of BLRs ($i_{\rm BLR}$) from the MCMC 
fittings of the QSO broad lines by a simple BLR model and estimating the 
orbital inclination angles ($i_{\rm orb}$) of the SMBBH orbital planes based 
on the DB hypothesis, we find that all the BLRs of QSOs in the DB sample are 
viewed from face-on orientations, while the BBH orbits viewed in the face-on 
and edge-on orientations are half-to-half. For the 21 QSOs in the DB sample, 
10 of them may have the BLRs aligned with the BBH orbital planes, while the 
others are not. However, after comparing the FWHM distributions and stacked 
\mgii\ profiles of these subsamples, no significant differences are found 
either between those obtained from the face-on and edge-on subsamples or 
between those from the aligned and misaligned subsamples. This further 
enhances the challenge on using Doppler boost hypothesis to interpret the 
optical periodicity of the systems in the DB sample. Even if the inclination 
angles of the BBH orbital planes are not correlated with the BLRs, it is still 
surprising that no single BLR is viewed at edge-on orientation, while both 
face-on and edge-on orientations of the BBH orbital planes are found in the DB 
sample with 21 QSOs.

\section*{Acknowledgements}
This work is supported by the National Key Program for Science and Technology
Research and Development (Grant No. 2016YFA0400704), the National Natural 
Science Foundation of China (NSFC) under grant number 11690024 and 11873056, 
and the Strategic Priority Program of the Chinese Academy of Sciences (Grant 
No. XDB 23040100).


%

\end{document}